\documentclass[prl,showpacs,twocolumn]{revtex4}

\usepackage{graphicx}
\usepackage{dcolumn}
\usepackage{bm}
\usepackage{amsmath}

\newcommand{\sba}{\begin{subeqnarray}}
\newcommand{\sea}{\end{subeqnarray}}

\def\cm-1{cm$^{-1}$}

\begin{document}

\title{A novel diagrammatic technique for the
       single-site Anderson model}
\author{V.\ A.\ Moskalenko$^{1,2}$}
\email{moskalen@thsun1.jinr.ru}
\author{P.\ Entel$^{3}$}
\author{D.\ F.\ Digor$^{1}$}
\author{L.\ A.\ Dohotaru$^{4}$}
\author{R.\ Citro$^{5}$}
\affiliation{$^{1}$Institute of Applied Physics, Moldova Academy
             of Sciences, Chisinau 2028, Moldova}
\affiliation{$^{2}$BLTP, Joint Institute for Nuclear Research,
             141980 Dubna, Russia}
\affiliation{$^{3}$University of Duisburg-Essen, 47048 Duisburg,
             Germany}
\affiliation{$^{4}$Technical University, Chisinau 2004,
             Moldova}
\affiliation{$^{5}$Dipartimento di Fisica E.\ R.\ Caianiello,
             Universit\'{a} degli Studi di Salerno and CNISM,
             Unit\'{a} di ricerca di Salerno, Via S. Allende,
             84081 Baronissi (SA), Italy}

\date{\today}

\begin{abstract}
%
A diagrammatic theory around the atomic limit is proposed for the
single-impurity Anderson model in which the strongly correlated
impurity electrons hybridize with free (uncorrelated) conduction
electrons. Using this diagrammatic approach, we prove the
existence of a linked cluster theorem for the vacuum diagrams and
derive Dyson type of equations for the localized and conduction
electrons and corresponding equations for the mixed propagators.
The system of equations can be closed by summing the infinite
series of ladder diagrams containing irreducible Green's
functions. The result allows to discuss resonances associated with
the quantum transitions at the impurity site.
%
\end{abstract}
%


\maketitle

The study of strongly-correlated electron systems has become
one of the most active fields of condensed matter physics
during the last decade. The properties of these systems cannot
be described by Fermi liquid theory. One of the important models
of strongly correlated electrons is the single-site or impurity model
introduced by Anderson in 1961 \cite{Anderson-1961}, which has been
intensively discussed in many papers. It is a model for a system of
free conduction electrons that interact with the system of a local spin
of an electron in the $d$ or $f$ shells of an impurity atom.
The impurity electrons are strongly correlated
because of the strong on-site Coulomb repulsion and interact with
the conduction electrons via exchange and hybridization.
Most previous work is based on the method of equations of motion for
retarded and advanced Green's functions and a truncating procedure
as proposed, for example, by Bogoliubov and Tjablikov
\cite{Bogoliubov-1959}. A first attempt to develop a diagrammatic
theory for this problem was undertaken by Barabanov in
\cite{Barabanov-1974}. With introduction of the dynamical mean field
theory, the interest in the Anderson impurity model has considerably
increased because infinite dimensional lattice models like the Hubbard model
can be mapped onto effective impurity models and a self-consistency
condition \cite{Georges-1996,Kotliar-2004}.

The Hamiltonian of the model taking only $s$-like electrons into
account is written as
%
\begin{eqnarray}
%
H & = & H_{0}+H_{int}, \quad H_{0} = H_{0}^{c} + H_{0}^{f},
\nonumber \\
H_{0}^{c} & = & \sum\limits_{\mathbf{k}\sigma }\epsilon
  (\mathbf{k})\ C_{\mathbf{k}\sigma}^{+} C_{\mathbf{k}\sigma},
  \; \; H_{0}^{f} = \epsilon_{f}\sum\limits_{\sigma }
                    f_{\sigma}^{+}f_{\sigma}
        + U n_{\uparrow}^{f}n_{\downarrow}^{f},
\nonumber \\
H_{int} & = & \frac{1}{\sqrt{N}}\sum\limits_{\mathbf{k}\sigma }
   \left( V_{ \mathbf{k}\sigma}f_{\sigma}^{+}C_{\mathbf{k}\sigma}
  + V_{\mathbf{k}\sigma}^{\ast}C_{\mathbf{k}\sigma}^{+}f_{\sigma}
   \right) ,
\label{1}
%
\end{eqnarray}
%
where
$n_{\sigma }^{f} = f_{\sigma }^{+}f_{\sigma }$,
$C_{\mathbf{k}\sigma }(C_{\mathbf{k}\sigma }^{+})$ and
$f_{\sigma }(f_{\sigma }^{+})$ are the annihilation (creation) operators
of conduction and impurity electrons with spin $\sigma$,
respectively; $\epsilon(\mathbf{k})$ is the kinetic energy of
the conduction band states $(\mathbf{k},\sigma )$; $\epsilon_{f}$
is the local energy of $f$-electrons and $N$ is the number
of lattice sites. $H_{int}$ is the hybridization interaction
between conduction and localized electrons. Summation over
$\mathbf{k}$  will be changed to an integral over the energy
$\epsilon(\mathbf{k})$ using the density of states
$\rho_{0}(\epsilon)$ of conduction electrons.

The term in the Hamiltonian involving $U$ describes the on-site
Coulomb interaction between two impurity electrons. This term is far
too large to be treated by perturbation theory. Therefore, we include
it in the non-interacting Hamiltonian $H_0$. Since this term
invalidates Wick's theorem for local electrons, we first of all
have to formulate the generalized Wick's theorem (GWT) for local
electrons, preserving the ordinary Wick theorem for conduction
electrons. Our GWT can be considered as the prescription which allows
to determine the irreducible Green's functions or Kubo cumulants.
A similar prescription has been used when discussing the properties
of the single-band Hubbard model
\cite{Vladimir-1990,Vakaru-1990,Moskalenko-1999}.

In interaction representation, the renormalized (Matsubara) Green's
functions of conduction and impurity electrons have the form:
%
\begin{eqnarray}
%
G(\mathbf{k,}\sigma,\tau \mid \mathbf{k}^{\prime},\sigma,^{\prime }
    \tau^{\prime})
  = - \left\langle \mathrm{T} C_{\mathbf{k}\sigma}(\tau )
      \overline{C}_{\mathbf{k}^{\prime}\sigma^{\prime}}(\tau^{\prime})
      U(\beta) \right\rangle_{0}^{c},
\\
g(\sigma,\tau \mid \sigma^{\prime},\tau^{\prime})
  = - \left\langle \mathrm{T}f_{\sigma}(\tau)
      \overline{f}_{\sigma^{\prime}}(\tau^{\prime})
      U(\beta ) \right\rangle_{0}^{c}.
\label{3}
%
\end{eqnarray}
%
Here $\tau$ and $\tau^{\prime}$ stand for the imaginary time with
$0 < \tau < \beta$ ($\beta$ is the inverse temperature) and
$\mathrm{T}$ is the chronological time ordering operator. The evolution
operator $U(\beta)$ is determined by the hybridization interaction $H_{int}$.
The statistical averaging in Eqs.\ (2) and (3) is carried out with
respect to the zero-order density matrix of the conduction and impurity
electrons.

The thermodynamic perturbation theory with respect to $H_{int}$
requires an appropriate generalization scheme for the evaluation of
statistical averages of the $\mathrm{T}$-products of localized
$f$-electron operators.

In zero-order approximation (negelecting the hybridization term),
the corresponding Green's functions have the form
$(\omega \equiv \omega_{n} =(2n+1)\pi /\beta)$:
%
\begin{eqnarray}
%
G_{\sigma \sigma^{\prime}}^{0}(\mathbf{k},\mathbf{k}^{\prime}
    \mid i\omega)
  = \delta_{\sigma\sigma^{\prime}} \delta_{\mathbf{kk}^{\prime}} /
    \left\{ i\omega - \epsilon(\mathbf{k)} \right\},
\nonumber \\
g_{\sigma \sigma^{\prime}}^0
  = \delta_{\sigma\sigma^{\prime}} g_{\sigma}^{0}(i\omega)
  = \delta_{\sigma\sigma^{\prime}}
    \left\{ \frac{1 - n_{\overline{\sigma}}}
                 {\lambda_{\sigma }(i\omega)}
          + \frac{n_{\overline{\sigma}}}
            {\overline{\lambda}_{\overline{\sigma}}(i\omega)}
    \right\},
\label{9}
%
\end{eqnarray}
%
where $(\overline{\sigma} = -\sigma)$ and
%
\begin{eqnarray}
%
\lambda_{\sigma}(i\omega) & = & i\omega + E_{0} - E_{\sigma},
    \quad
    \overline{\lambda}_{\overline{\sigma}}(i\omega)
  = i\omega + E_{\overline{\sigma}} - E_{2},
\nonumber \\
Z_{0} & = & e^{-\beta E_{0}} + e^{-\beta E_{\sigma}}
  + e^{-\beta E_{\overline{\sigma}}}
  + e^{-\beta E_{2}},
\nonumber \\
n_{\overline{\sigma}} & = &\left\{
    e^{-\beta E_{\overline{\sigma}}} + e^{-\beta E_{2}}
    \right\} / Z_{0}.
\nonumber
%
\end{eqnarray}
%
In the case of $f$-electrons, we can formulate the identity
corresponding to the GWT:
%
\begin{eqnarray}
%
\left\langle
\mathrm{T}f_{1}f_{2}\overline{f}_{3}\overline{f}_{4}\right\rangle_{0} =
  \left\langle \mathrm{T}f_{1}\overline{f}_{4}\right\rangle_{0}
  \left\langle \mathrm{T}f_{2} \overline{f}_{3}\right\rangle_{0}
\nonumber \\ {}
 - \left\langle \mathrm{T}f_{1}\overline{f}_{3}\right\rangle_{0}
   \left\langle \mathrm{T}f_{2}\overline{f}_{4}\right\rangle_{0}
 + \left\langle \mathrm{T}f_{1}f_{2}\overline{f}_{3}\overline{f}_{4}
   \right\rangle_{0}^{ir},
\label{10}
%
\end{eqnarray}
%
or
%
\begin{eqnarray}
%
g_{2}^{0}(1,2|3,4) & = & g^{0}(1|4)g^{0}(2|3)
\nonumber \\
& - & g^{0}(1|3)g^{0}(2|4) + g_{2}^{(0)ir}(1,2|3,4),
\label{11}
%
\end{eqnarray}
%
where $n$ stands for $(\sigma_{n},\tau_{n})$. The generalization
for more complicated averages of type \newline
$g_{n}^{0}(1,...,n\mid n+1,...,2n)
 = (-1)^{n}\left\langle \mathrm{T} f_{1}...f_{n}
   \overline{f}_{n+1}...\overline{f}_{2n} \right\rangle_{0}$
is straightforward, i.e., the right-hand part of this
relation will contain $n!$ terms of ordinary Wick type (chain
diagrams) and also the different products of irreducible functions
with the same total number of operators. These definitions are the
simplification of the corresponding ones for Hubbard and other
lattice models proposed eaarlier by us. The calculation of the
simplest irreducible function like, for example,
$g_{2}^{(0)ir}(12|34)$ is rather cumbersome
but straightforward. Then, we  determine the Fourier representation
of $g_{2}^{(0)ir}[\sigma_{1},\tau_{1};\sigma_{2},
   \tau_{2}|\sigma_{3},\tau_{3};\sigma_{4},\tau_{4}]$,
for which exists spin and frequency  conservation,
$\sigma_{1} + \sigma_{2} = \sigma_{3} + \sigma_{4}$,
$\omega_{1} + \omega_{2} - \omega_{3} - \omega_{4} = 0$,
Thus, we dispose of the rules to deal with
chronological averages in thermodynamic perturbation theory.

First of all, we determine the thermodynamic potential $F$ of
the system,
%
\begin{eqnarray}
%
F = F_{0} - \frac{1}{\beta} \ln
    \left\langle U(\beta) \right\rangle_{0},
\label{7}
%
\end{eqnarray}
%
where $F_0$ refers to the free impurity atom and conduction
electrons. The diagrams which determine the thermodynamic
potential have no external lines and represent the vacuum.
%
\begin{figure}[!h]
%
\centering
\includegraphics[width=0.5\textwidth,clip]{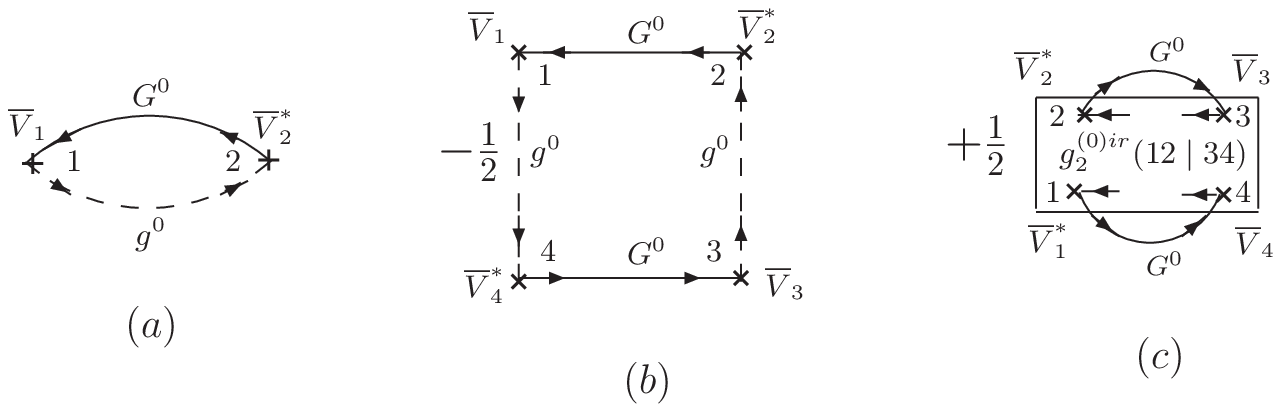}
\caption{The simplest connected vacuum diagrams of the normal state.
The diagram $(a)$ is of second and $(b)$ and $(c)$ are of fourth order
in the hybridization interaction. Here,
$\overline{V}_{n} = V_{n}/\sqrt{N}$.}
\label{fig-1}
\end{figure}
%
Figure \ref{fig-1} shows the simplest connected vacuum
diagrams of the normal (non-superconducting) state with
the process of hybridization of $C$- and $f$-electrons. The zero
order propagators of conduction and impurity electrons are
represented by solid and dashed lines, respectively. These
lines connect the crosses which depict the impurity states. To
each of the crosses two arrows are attached, one of which is
ingoing and the other outgoing. They stand for annihilation and
creation of electrons. The index $n$ denotes $(\sigma_{n},\tau_{n})$
for impurity and $(\mathbf{k}_{n},\sigma _{n},\tau_{n})$ for
conduction electrons. The rectangles with $2n$ crosses represent the
irreducible $g_{n}^{(0)ir}$ Green's function, $g_{n}^{(0)ir}$.

Besides the vacuum diagrams of fourth order in Fig.\ \ref{fig-1} $(b)$
and $(c)$, there is also one disconnected diagram composed of two
diagrams of type $(a)$ containing an additional factor $1/2!$.
This is repeated in high order of perturbation theory and permits us to
formulate the linked cluster theorem, which has the form
%
\begin{equation}
%
\left\langle U(\beta) \right\rangle_{0}
 = \exp \left\langle U(\beta) \right\rangle_{0}^{c},
\label{15}
%
\end{equation}
%
where $\left\langle U(\beta) \right\rangle_{0}^{c}$ contains only
connected diagrams and is equal to zero when hybridization is
swichted off.

\begin{figure}[!h]
%
\centering
\includegraphics[width=0.5\textwidth,clip]{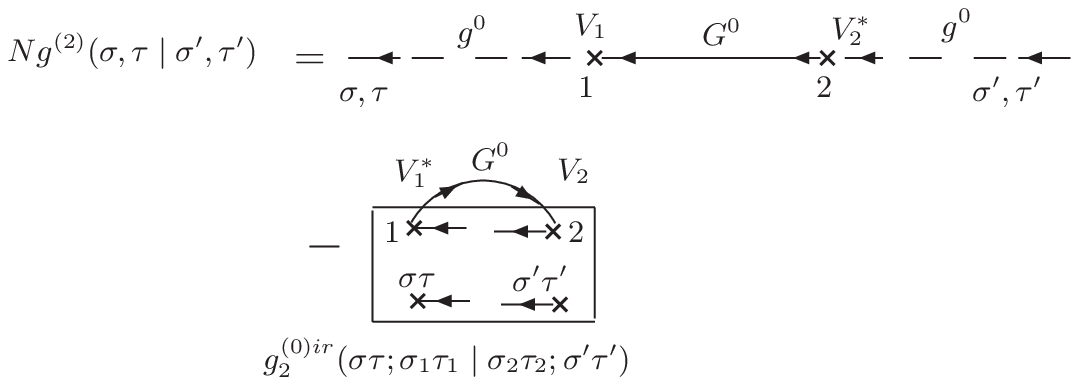}
\caption{The second order perturbation contribution
to the impurity electron propagator.}
\label{fig-2}
%
\end{figure}

Now, we consider the diagrammatical analysis of the perturbation
series for the renormalized propagator in Eq.\ (3). The simplest
contributions to $g$ in second order are represented in
Fig.\ \ref{fig-2}. All diagrams contain two external points with
attached arrows fixed by the arguments of the Green's functions.
At the inner points of the diagrams a summation over
$\sigma_{n},\mathbf{k}_{n}$ and integration over $\tau_{n}$
is included. The new diagrammatical elements which appear, are the
irreducible two-particle Green's functions. The renormalization
of these quantities is not considered here, only
the renormalization of the propagators due to hybridization.

The final equations for the renormalized Green's functions are more
conveniently written down in Fourier representation. For the
propagator of conduction electrons we obtain:
%
\begin{align}
%
G_{\sigma\sigma^{\prime}}(\mathbf{k},\mathbf{k}^{\prime}
   | \,i\omega )
& = \delta_{\mathbf{kk}^{\prime}}\delta_{\sigma\sigma^{\prime}}
   G_{\sigma}^{0}(\mathbf{k} | \,i\omega)
\nonumber \\ & {}
 + \frac{V_{\mathbf{k}}^{\ast}V_{\mathbf{k}^{\prime}}}{N}
   G_{\sigma}^{0}(\mathbf{k} | i\omega) \,
   g_{\sigma\sigma^{\prime}}(i\omega) \,
   G_{\sigma^{\prime}}^{0}(\mathbf{k}^{\prime} | \,i\omega).
\label{17}
%
\end{align}
%
This renormalized propagator is expressed by the full
propagator $g$ of the impurity electrons. Now, it is necessary to
obtain the corresponding equation for the full impurity electron
propagator. Because the subsystem of $f$-electrons is strongly
correlated, we have to introduce the correlation function
$Z_{\sigma\sigma^{\prime}}$ which is represented by strongly
connected diagrams containig irreducible Green's functions
\cite{Vladimir-1990,Vakaru-1990,Moskalenko-1999}
and
$\Lambda_{\sigma}(i\omega)=g^{0}_{\sigma}(i\omega)+Z_{\sigma}(i\omega)$.
To this we introduce new quantities:
%
\begin{align}
%
\frac{1}{N}  &  \sum\limits_{\mathbf{k}_{1}\mathbf{k}_{2}}
   V_{\mathbf{k}_{2}}^{\ast}V_{\mathbf{k}_{1}}
   G^{0}(\mathbf{k}_{1},\sigma_{1},\tau_{1} |
   \mathbf{k}_{2},\sigma_{2},\tau_{2})
 = \frac{1}{N}\sum\limits_{\mathbf{k}_{1}}|V_{\mathbf{k}_{1}}|^{2}
\nonumber \\ & {} \times
   G_{\sigma_{1}\sigma_{2}}^{0}(\mathbf{k}|\tau_{1}
 - \tau_{2}) \equiv \delta_{\sigma_{1}\sigma_{2}}
   G_{\sigma_{1}}^{0}(\tau_{1} - \tau_{2}),
\label{10}
%
\end{align}
%
which allow to simplify the structure of equation for the
$f$-electron propagator, which can be written as
%
\begin{eqnarray}
%
g_{\sigma }(i\omega ) & = & \big \lbrace
   \Lambda_{\sigma}(i\omega) 
\nonumber \\
&& 
{}
 - G_{\overline{\sigma}}^{0}(-i\omega)
   \Lambda_{\sigma}(i\omega)\Lambda_{\overline{\sigma}}(-i\omega)
   \big \rbrace / d_{\sigma }(i\omega ).
\\
d_{\sigma}(i\omega) & = & [1 - \Lambda_{\sigma}(i\omega)
   G_{\sigma}^{0}(i\omega)][1 - \Lambda_{\overline{\sigma}}(-i\omega)
   G_{\overline{\sigma}}^{0}(-i\omega)].
\nonumber
%
\end{eqnarray}
%
Equation (9) has originally been derived by Anderson
\cite{Anderson-1961} by using the equation of motion method in
which the propagator $g_{\sigma}(i\omega )$ corresponds to the
$t$-matrix for non-spin-flip scattering.

\begin{figure}[!h]
%
\centering
\includegraphics[width=0.45\textwidth,clip]{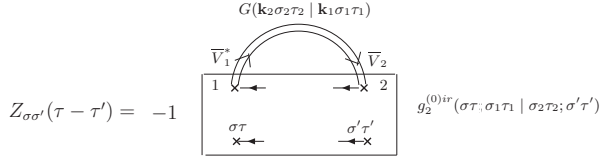}
\caption{Diagrammatic representation of the approximative scheme
used in the evaluation of the correlated function. The solid
double lines with arrows depict the renormalized one-particle
Green's function of conduction electrons. The rectangle depicts
the irreducible Green's function of impurity electrons.}
 \label{fig-3}
%
\end{figure}

The main approximation which we employ here corresponds to the
ladder approximation (which contain the main contributions of spin
and charge fluctuations) and is schematically shown
in Fig.\,\ref{fig-3}.
%
%
We investigate the properties of the normal state of the
single-site Anderson model by using Eqs.\ (9) and (11) for the
corresponding Green's functions of conduction and
impurity electrons. In the ladder approximation only the
simplest irreducible Green's function $g_{2}^{(0)ir}$ is
iterated many times. In the paramagnetic phase
($\sigma^{\prime} = \sigma$) the equation for
$Z$ reduces to
%
\begin{align}
%
Z_{\sigma }(i\omega) & =
  -\frac{1}{\beta}\sum\limits_{\omega_{1}}
   \sum\limits_{\sigma_{1}}G_{\sigma_{1}}(i\omega_{1})
\nonumber \\ & {} \times
  \widetilde{g}_{2}^{(0)ir}
  [\sigma,i\omega;\sigma_{1},i\omega_{1} |
   \sigma_{1},i\omega_{1};\sigma,i\omega],
 \label{7}
%
\end{align}
where
%
\begin{equation}
%
G_{\sigma }(i\omega) =
   \frac{1}{N}\sum\limits_{\mathbf{k}_{1}\mathbf{k}_{2}}
   V_{\mathbf{k}_{1}}^{\ast }V_{\mathbf{k}_{2}}
   G_{\sigma}(\mathbf{k}_{1},\mathbf{k}_{2}|i\omega).
\label{8}
%
\end{equation}
%
Using Eqs.\ (9) and (11) $G$ can be written as
%
\begin{align}
%
G_{\sigma }(i\omega ) & = G_{\sigma}^{0}(i\omega )
 + [G_{\sigma}^{0}(i\omega)]^{2}g_{\sigma}(i\omega)
\nonumber \\ &
 = \frac{G_{\sigma}^{0}(i\omega)}{1-\Lambda_{\sigma}(i\omega)
   G_{\sigma}^{0}(i\omega)}.
\label{12}
%
\end{align}
%
Then, using the definition of the correlation function,
$\Lambda_{\sigma}(i\omega)$, and approximation (12) for
$Z_{\sigma }(i\omega )$, we obtain the final integral
equation for $\Lambda_{\sigma}$:
%
\begin{eqnarray}
%
\Lambda_{\sigma}(i\omega) & = &
   g_{\sigma}^{(0)}(i\omega) - \frac{1}{\beta}
   \sum\limits_{\omega_{1}}\sum\limits_{\sigma_{1}}
   \frac{G_{\sigma_{1}}^{0}(i\omega_{1})}
   {1 - \Lambda_{\sigma_{1}}(i\omega_{1})
   G_{\sigma_{1}}^{0}(i\omega_{1})}
\nonumber \\ && {} \times
   \widetilde{g}_{2}^{(0)ir}[\sigma,i\omega;\sigma_{1},i\omega_{1} |
   \sigma_{1},i\omega_{1};\sigma,i\omega].
\label{10}
%
\end{eqnarray}
%

For the symmetric case when $\epsilon_{f} = -U/2 < 0$ and
$\epsilon_{f} + U = U/2 >0$, we have more simple equations:
%
\begin{eqnarray}
%
g_{\sigma}^{0}(i\omega) = i\omega
   \left\{(i\omega)^{2} - (U/2)^{2}\right\},
\nonumber \\
Z_{0} = 2(1 + \exp(\beta U/2)), \quad n_{\sigma} = 1/2
%
\end{eqnarray}
%
together with the antisymmetry property of the zero-order impurity
Green's function, $g_{\sigma}^{0}(-i\omega) =
-g_{\sigma}^{0}(i\omega)$. In addition, if we assume that the
matrix element $V(\epsilon)$ and the bare density of states,
$\rho_{0}(\epsilon)$ are even function of its argument, then
$G_{\sigma}^{0}(i\omega)$ is also antisymmetric,
$G_{\sigma}^{0}(-i\omega) = -G_{\sigma}^{0}(i\omega)$. This allows
us to search for the antisymmetric solution,
%
\begin{equation}
%
\Lambda_{\sigma}(-i\omega) = -\Lambda_{\sigma}(i\omega) ,
\label{37}
%
\end{equation}
%
of Eq.\ (15). The analytical continuation of $G_{\sigma}^{0}(i\omega)$
has the form
%
\begin{align}
%
G_{\sigma}^{0}(E+i\delta) & = I(E) - i\Gamma (E),
\nonumber \\
I(E) & = \mathrm{P} \! \int \! d\epsilon \, \rho_{0}(\epsilon)
   \frac{|V(\epsilon)|^{2}}{E-\epsilon},
\nonumber \\
   \Gamma (E) & = \pi \rho_{0}(E) |V(E)|^{2},
\label{39}
%
\end{align}
%
where $I(E)$ is the principal part of the integral. This function is
antisymmetric. $\Gamma (E)$ is the band width of the virtual level
and an even function of the energy. The symmetric impurity Anderson
model is easier to treat since it leads to a simple form for the
irreducible two-particle Green's functions in the different spin and
frequency channels yielding
%
\begin{align}
%
\Lambda_{\sigma}(i\omega) &
 = \frac{3(U/2)^{2}G_{\sigma}^{0}(i\omega)}
   {[(i\omega)^{2} - (U/2)^{2}]^2
    [1 - \Lambda_{\sigma}(i\omega)G_{\sigma}^{0}(i\omega)]}
\nonumber \\ & {}
 + \frac{i\omega}{(i\omega)^{2} - (U/2)^{2}},
 \label{19}
%
\end{align}
%
where we have used results obtained for the one-band Hubbard model
\cite{Vladimir-1990,Vakaru-1990,Moskalenko-1999}.
There are two solutions of Eq.\ (19). The physical one with the
correct asymptotic behavior,
$\Lambda_{\sigma}(i\omega)\rightarrow 1/i\omega$ when $ |\omega|$ tends
to infinity, has the form:
%
\begin{align}
%
\Lambda_{\sigma}(i\omega) & =
   \frac{1}{2G_{\sigma}^{0}(i\omega)
   [(i\omega)^{2} - (U/2)^{2}]}
\nonumber \\ & {} \times
   \Big \lbrace
   \left[ (i\omega )^{2}-(U/2)^{2}
+ i\omega G_{\sigma}^{0}(i\omega) \right]
\nonumber \\ & {}
- \left[ (i\omega)^2 - (U/2)^2 - i\omega G_{\sigma}^{0}(i\omega) \right]
  \sqrt{1 - 12Q(i\omega)}
   \Big \rbrace ,
\label{41}
%
\end{align}
%
where
%
\begin{equation}
%
Q(i\omega ) = \left(
  \frac{(U/2) G_{\sigma}^{0}(i\omega)}
  {(i\omega)^{2} - (U/2)^{2} - i\omega G_{\sigma}^{0}(i\omega)}
  \right)^{\! 2}.
\label{42}
%
\end{equation}
%
Finally, we obtain for the renormalized impurity electron propagator:

%
\begin{align}
%
g(i\omega) = 2 & \Big \lbrace
   g^{0}(i\omega)/G^{0}(i\omega)
 + 3Q(i\omega) [ g^{0}(i\omega) 
\nonumber \\ & {}
   - 1/G^{0}(i\omega)]^{2}  \Big \rbrace \Big \lbrace
     [1/G^{0}(i\omega)
\nonumber \\ & {}
 - g^{0}(i\omega)][1 + \sqrt{1 - 12Q(i\omega)}
\nonumber \\ & {}
 + 6Q(i\omega) \lbrace g^{0}(i\omega)G^{0}(i\omega) -1 \rbrace]
   \Big \rbrace^{-1}.
\label{43}
%
\end{align}
%
From the analytical continuation we obtain the spectral function of
impurity electrons, which is given in Hubbard I approximation by
%
\begin{align}
%
A_{f}^{I}(E) = \frac{2\Gamma (E)}
    {[E - (U/2)^{2}/E - I(E)]^{2} + \Gamma^{2}(E)}.
\label{25}
%
\end{align}
%
Equation (23) has two resonances at $E = \pm E_{0}$, where $E_{0}$
is determined from the condition that the brachet in the
denominotor of (23) is equal to zero. These resonances are
broadened by $\Gamma(E)$ which determines height and width of both
resonances. This approximation yields a vanishing spectral
function on the Fermi surface, where $E = 0$. Only beyond Hubbard
I approximation, when strong correlations are taken into account,
we obtain from Eq.\ (20) the spectral function of impurity
electrons in the vicinity of the Fermi level having the following
form,
%
\begin{equation}
%
A_{f}(E) \simeq  \frac{12 (2\Gamma/U)^{2} \,
   /\Gamma}
   {1 + 6(2\Gamma/U)^{2} +
   \sqrt{1 + 12 (2\Gamma/U)^{2}}} .
%
\end{equation}
%
%
\begin{figure}[!h]
%
\centering
\includegraphics[width=0.45\textwidth,clip]{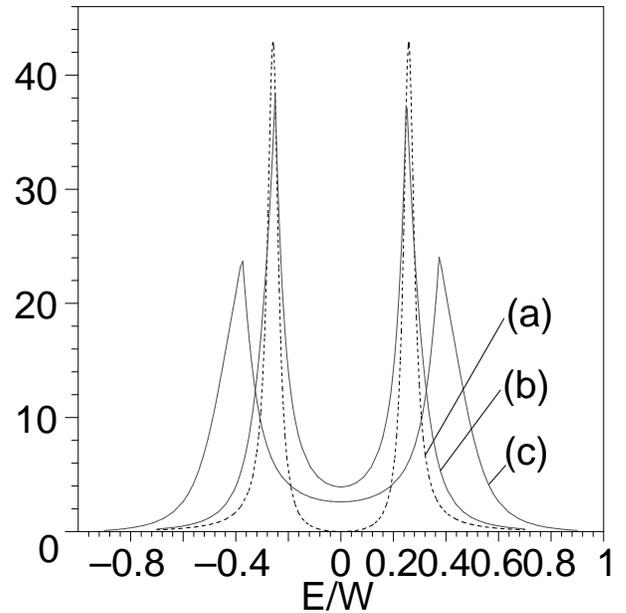}
\caption{Spectral function $A_{f}(E)\times W$ for different values
of the theory parameters as function of energy $E/W$ in the
Hubbard I approximation (case (a)) and in the our ladder
approximation (cases (b) and (c)). In the cases (a) and (b)
$U=1\,eV,\quad W=2\,eV,\quad \Gamma =0.1\,eV $ and in case case
(c) $U=1.5\,eV,\quad W=2\,eV,\quad \Gamma =0.15\,eV $. }
\label{fig-4}
%
\end{figure}
%
\vspace{-5mm}

In conclusion we state that our diagrammitc approach allows to
describe in ladder approximation for the irreducible Green's
function two resonances  at energies $E = \pm E_{0}$ and the
feature at $E=0$ given by Eq.\ (24). The distance between two
peaks (Figure \ref{fig-4}) is determined by the value of $U$ but
the height and the width of the peaks by $\Gamma$. Beyond the
Hubbard I approximation the non zero values of spectral function
appear for $E=0$.

This work has been supported by a Grant from the
Heisenberg-Landau Program. V.\ A.\ M. would like to thank the Universities
of Duisburg-Essen and Salerno for hospitality and financial support.

%



\begin{thebibliography}{99}
%

\bibitem{Anderson-1961} 
   P.\ W.\ Anderson, Phys.\ Rev.\ \textbf{124}, 41 (1961).

\bibitem{Bogoliubov-1959} 
   N.\ N.\ Bogoliubov and S.\ V.\ Tjablikov, Doklady AN
   USSR,\ \textbf{126}, 53 (1959) [in Russian].

\bibitem{Barabanov-1974} 
   A.\ F.\ Barabanov, C.\ A.\ Kikoin and L.\ A.\ Maximov,
   Theor.\ Math.\ Phys.\ \textbf{20}, 364 (1974).

\bibitem{Georges-1996} 
   A.\ Georges, G.\ Kotliar, W.\ Krauth and M.\ J.\
   Rozenberg, Rev.\ Mod.\ Phys.\ \textbf{8}, 13 (1996).

\bibitem{Kotliar-2004} 
   G.\ Kotliar and D.\ Vollhardt, Physics Today \textbf{57}, 53 (2004).

\bibitem{Vladimir-1990} 
   M.\ I.\ Vladimir and V.\ A.\ Moskalenko, Theor.\ Math.\
   Phys.\ \textbf{82}, 301 (1990).

\bibitem{Vakaru-1990} 
   S.\ I.\ Vakaru, M.\ I.\ Vladimir and V.\ A.\ Moskalenko,
   Theor.\ Math.\ Phys.\ \textbf{85}, 1185 (1990).

\bibitem{Moskalenko-1999} 
   V.\ A.\ Moskalenko, P.\ Entel and D.\ F.\ Digor, Phys.\
   Rev.\ B \textbf{59}, 619 (1999).

\end{thebibliography}
\end{document}